# Studies of "Kapustinsky's" light pulser timing characteristics

B.K.Lubsandorzhiev, Y.E.Vyatchin

*Institute for Nuclear Research RAS*

Corresponding author:

Tel: +7(095)1353161; fax: +7(095)1352268;

e-mail: lubsand@pcbai10.inr.ruhep.ru

Address: 117312 Moscow Russia, pr-t 60-letiya Oktyabrya 7A

Institute for Nuclear Research RAS

We present the results of studies of a nanosecond light pulser built following *J.S.Kapustinsky et al* original design and using bright InGaN/GaN ultraviolet and blue LEDs produced by NICHIA CHEMICAL. It is shown how timing characteristics of the pulser depend on the type of LED and the value of power supply voltage.

In the 1985 J.S.Kapustinsky and his colleagues published a beautiful scheme of an inexpensive compact nanosecond LED pulser [1]. Since that time the pulser has become very popular in astroparticle physics experiments at least where it's widely used for time and amplitude calibrations: in high energy neutrino telescopes like NT-200 in Lake Baikal [2] and ANTARES in the Middeterrenian Sea [3], the imaging atmospheric Cherenkov telescope H.E.S.S. [4], the extremely high energy cosmic ray detector AUGER [5] etc. The popularity ensued from the pulser's high performances, simplicity, convenience, and robustness.

The pulser is based on a fast discharge of a small capacitor via a complementary pair of RF transistors. An electrical scheme of the pulser is shown in Fig.1. In the present work we used a blue and UV LEDs, NICHIA NSPB300A and NSHU590E respectively, as a light source in the pulser. Emission spectra of the LEDs have peaks occurring at 470 nm and 370 nm respectively [6]. The blue LED was out of an "old" batch of LEDs manufactured in 1996. In our previous paper [7] it was demonstrated that the old LEDs are substantially faster in contrary to new LEDs of the same type presently available in the market. As to the UV LEDs so far no differences have been seen in the emission kinetics of the old and new LEDs.

One of doubtless advantages of the pulser is a possibility to adjust quite easily the light pulse intensity of LEDs by varying a power supply voltage in contrary to LED pulsers based on avalanche transistors where to change amplitude of light pulses in a wide range is more complicated. The advent of ultra bright InGaN/GaN LEDs at the beginning of 90s made the "Kapustinsky's" pulser much more brighter with up to $10^7 \div 10^8$ photons per pulse and ~2÷3 ns pulse width. It made the pulser very much attractive for use in many fields of experimental physics.

We measured light pulses temporal profiles and a relative change of a delay time of output light pulse from the pulser trigger pulse as functions of the power supply voltage. The pulser properties were studied with a set-up shown schematically in Fig.2. A fast ET9116B PMT [8] produced by Electron Tubes Ltd with ~0.1 ns time jitter was used for the measurements. The pulser was triggered by a positive pulse with an amplitude of 3V and 50 ns width from a fast pulse generator (Stanford DG535). A short optical cable (*OC*) was used to illuminate the PMT photocathode from the pulser. The PMT anode signals were amplified by a fast transimpedance preamplifier (*Preamp*) and LeCroy 612AM amplifier (*Amp*) and fed to a constant fraction discriminator (*CFD$_1$*). TDC LeCroy 2228A were used for the light pulses profiles measurements. An electronic jitter of the whole set-up is less than 50 ps. The power supply voltage $|V_{cc}|$ is varied in a range of 7V ÷24V. The lower limit corresponds to the voltage at which the light pulses become to be detected by the PMT. For the pulser's transistors safety we were forced to set 24 V as the upper limit of $|V_{cc}|$.

The light pulses waveforms were measured by a time correlated single photon counting technique [9]. In this method a PMT photocathode illumination is reduced to a single photoelectron level and if a PMT time jitter is small, measurements results will basically account for emission kinetics of a light source under studies. In our case as mentioned above the PMT time jitter is ~0.1 ns and inserts nearly negligible contribution to the LEDs light emission kinetics. LeCroy 2249A ADC was used to control a single photoelectron mode and to define a discriminator's threshold which was set to be a quarter of the mean value a single photoelectron charge.

Dependencies of the LEDs light pulses widths on the value of the pulser power supply voltage $|V_{cc}|$ are shown in Fig.3. The light pulse temporal profile of the UV LED has noticeably stronger dependence on $|V_{cc}|$ in comparison with the blue LED light emission kinetics. Furthermore the UV LED starts to emit light pulses at higher values of $|V_{cc}|$ due to higher current threshold of the LED. The width of light pulse (fwhm) widens with the increase of the power supply voltage $|V_{cc}|$ from 0.8 ns at 9 V to 2.8 ns at 24 V for the UV LED and from 1.7 ns at 7 V to 2.2 ns at 24 V for the blue LED. The light pulses waveforms of the UV LED for the power supply voltages of 9 V and 24 V are shown in Fig.4, dotted and full lines respectively. The second peaks clearly seen in both spectra are attributed to photoelectrons backscattered from the PMT's first dynode [10]. One can see that the waveform at 24 V is not only slower but conspicuously noisier in comparison with the 9 V case. It could be explained by the fact that at high current pulses the UV LED has slow emission component due likely to deep levels excitations [11,12]. This slow emission accounts for high background appearing at high values of $|V_{cc}|$. We don't show here similar waveforms for the blue LED because they are not so good distinguished as in case of the UV LED.

A relative change of the delay time between the pulser trigger pulse and output light pulse decreases slightly by ~480 ps with the increase of $|V_{cc}|$ in a range of 7÷24V for the blue LED. In contrary with the blue LED the delay time for the UV LED increases by ~730 ps with increase of $|V_{cc}|$ in a range of 9÷24 V.

The experimental data for the blue and UV LEDs are shown in Fig. 5 by filled circles and triangles respectively. To determine the delay time we measured a relative shift of the main peak position of the PMT photoelectron transit time distribution with the change of $|V_{cc}|$.

Thus the NICHIA blue and UV LEDs are considerably faster at lower values of the pulser power supply voltage $|V_{cc}|$. The fastest light emission kinetics has been observed with the UV LED at the lowest possible value of $|V_{cc}|$. In this case the light pulse width is about 0.8 ns and one can see how clearly the backscattered photoelectrons peak is resolved, see Fig.4. Furthermore the timing characteristics of the pulser based on the NICHIA UV LED have stronger dependence on the value of the pulser power supply voltage in contrary to the pulser based on the NICHIA blue LED.

The authors are indebted very much to Dr. E.Lorenz for initiating the present work and Dr. V.Ch.Lubsandorzhieva for careful reading the paper and many invaluable remarks. The work has been done partly due to support of Russian Foundation of Basic Research grant # 02-02-17365.

**Figures.**

Figure 1. Electrical scheme of the pulser.

Figure 2. Experimental set-up.

Figure 3. Width of LED light pulse as a function of power supply voltage $|V_{cc}|$: ● - blue LED; ▲ - UV LED.

Figure 4. Light pulse waveform of the UV LED at two power supply voltages $|V_{cc}|$: dotted line - 9 V; full line - 24 V.

Figure 5. Dependence of the delay time between light pulse and triggering pulse of the pulser on power supply voltage $|V_{cc}|$: ● - blue LED; ▲ - UV LED.

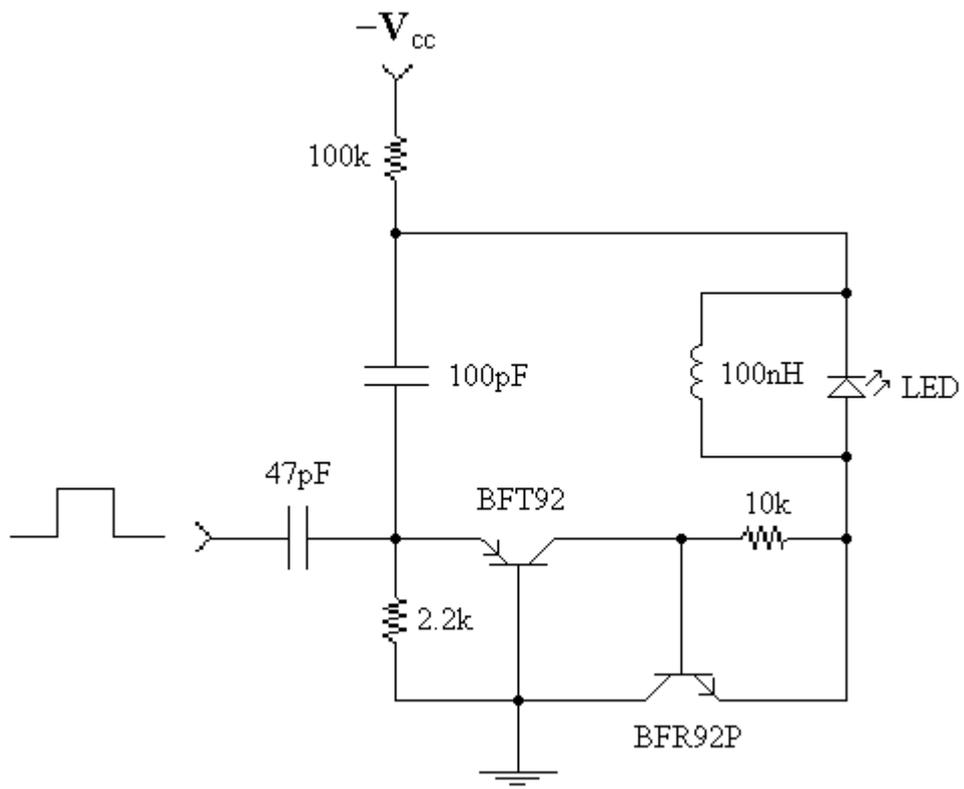

Fig.1.

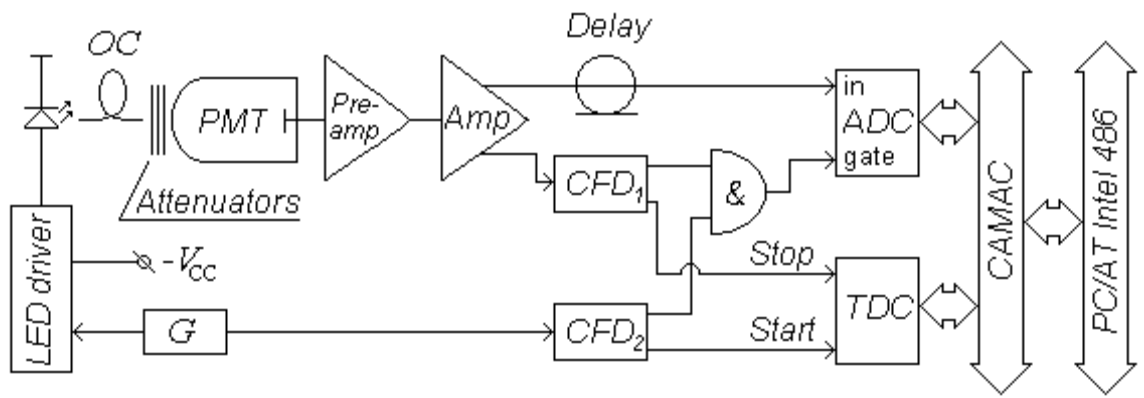

Fig.2

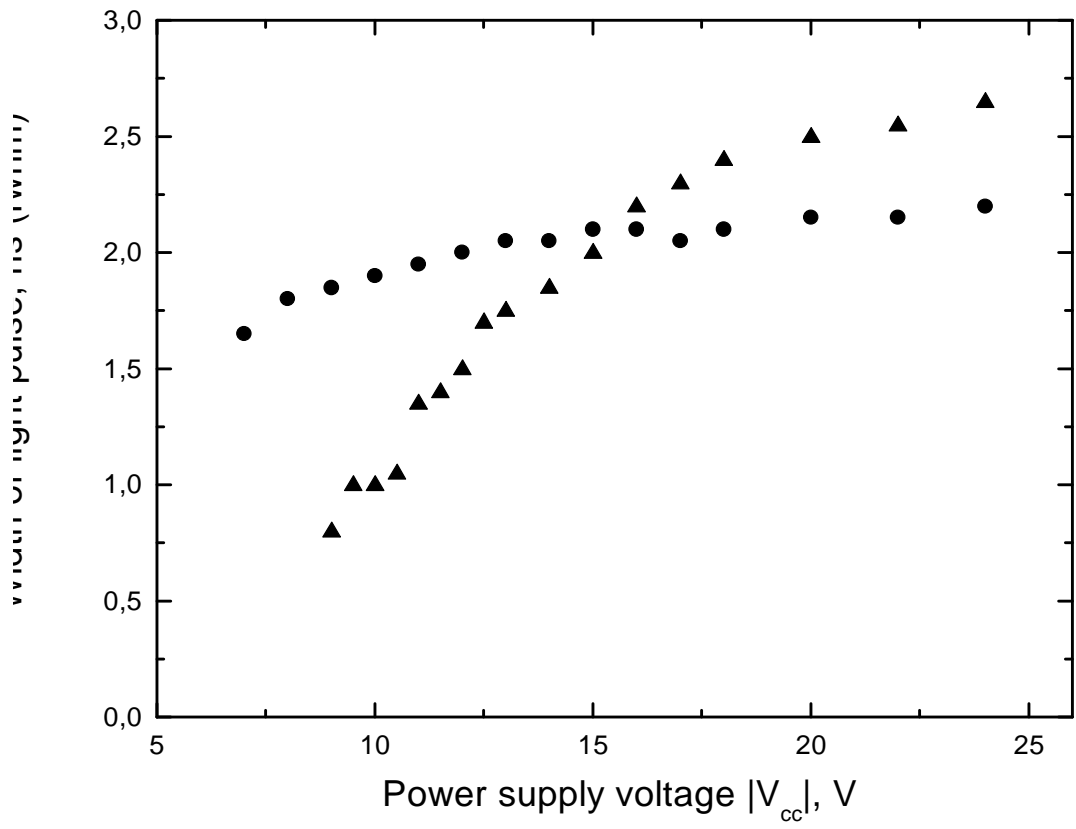

Fig.3.

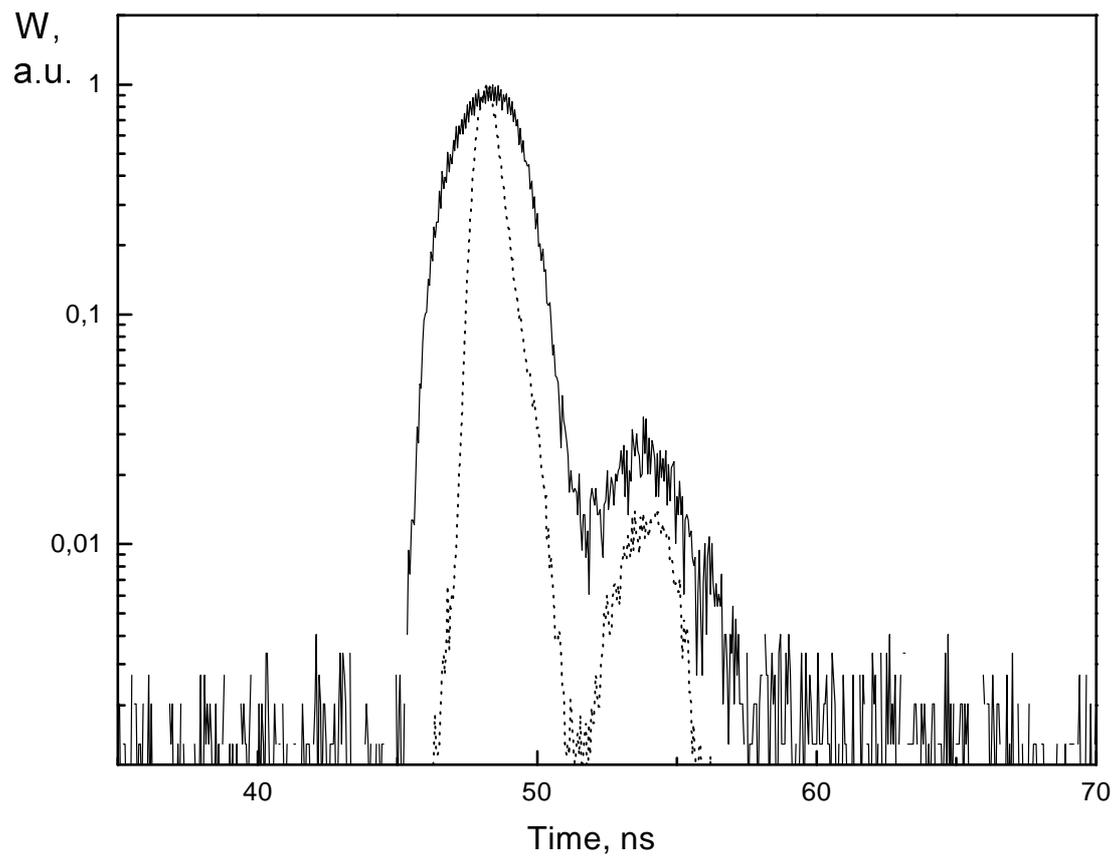

Fig.4.

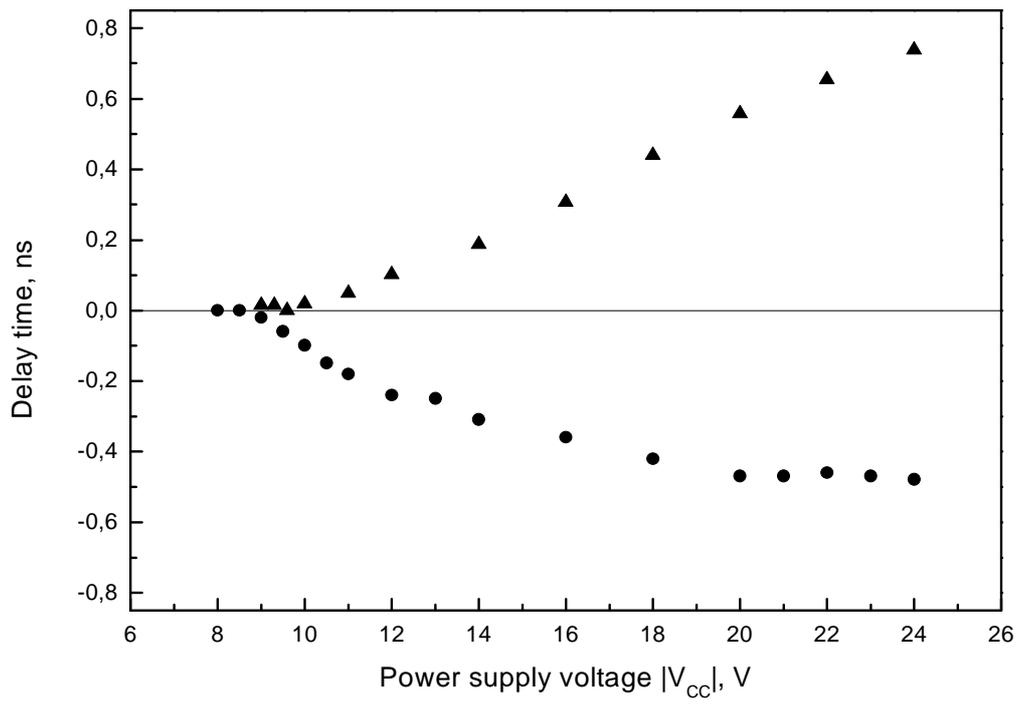

Fig.5.